\documentclass[11pt]{article}

\input epsf.sty

\def\lromn#1{\uppercase\expandafter{\romannumeral#1}}

\begin{document}
\begin{flushright}
\today \\
\end{flushright}

\begin{center}
\begin{Large}
\textbf{
Yet another symmetry breaking to be discovered
}
\end{Large}

\vspace{2cm}
M.~Yoshimura

\vspace{0.5cm}
Center of Quantum Universe, Faculty of
Science, Okayama University \\
Tsushima-naka 3-1-1 Kita-ku Okayama
700-8530 Japan
\end{center}

\vspace{3cm}

\begin{center}
\begin{Large}
{\bf ABSTRACT}
\end{Large}
\end{center}

The discovery of spontaneous symmetry breaking 
in particle physics was the greatest contribution
in Nambu's achievements.
There is another class of symmetries that
exist in the low energy nature, yet
is doomed to be broken at high energy,  due to a lack
of protection of the gauge symmetry.
I shall review our approach to
search for this class of symmetry breaking, the lepton number violation
 linked to generation of
the matter-antimatter asymmetry in our universe.

\newpage

\section
{\bf My old memory of a great physicist at Chicago}

Days I had frequently chances to speak to Yoichiro Nambu
started at December 1967 and ended at
August 1970, during which I was a graduate student
at University of Chicago.
Even before I went to Chicago,
Nambu's name was shining at
University of Tokyo 
where I spent 
undergraduate and 
two graduate years.
I did not know his famous 
superconductivity model with Jona-Lassinio, though.

The very first impression I had with
Nambu was his extreme politeness.
People often talk of his modesty,
but I felt more than that.
I was very much bewildered by his attitude,
since I had expected a different impression as a type of American professors.
Later I found that his attitude towards physicists
is invariant, irrespective of students, postdocs and
professors.

I decided to knock on the door of his office at least
once a week, whether or not
I have questions or something to report on.
I found that he would always welcome me and
started to talk of subjects he was interested in.
I must say that this was unusual, because
he frequently spoke of his latest ideas at that time prior
to publication.

One of the works Nambu was much involved around that time
was the  infinite-component field theory \cite{infinite component}.
He presumably tried to derive a mass formula for
all elementary particles.
This work followed his famous work with Han,
Han-Nambu model and three triplet model.
I studied three triplet model and found ideas fresh:
especially exchange of colored vector boson
to classify lowest energy hadrons was fascinating.
But Nambu somehow was not satisfied with those models and
he already embarked on the infinite-component
field theory.

At some time during those days
Streater and his young collaborator
published a no-go theorem to this
type of local infinite-component theory:
their claim was that this approach leads to
violation of  unitarity or causality.
Nambu immediately invited them and
a young guy gave a seminar, very short one
less than half an hour.
Although I did not understand this seminar,
Nambu immediately recognized this as a fatal block
against his approach (Gell-Mann  and many others were also
involved in the infinite-component at that time).
Thus, he very quickly abandoned his theory.

Around this time the Veneziano model and its extension,
the dual resonance model, were gaining a popularity.
Its particle content of infinitely many resonances
and high energy Regge behavior of scattering amplitudes
was also of Nambu's interest.
My opinion on Nambu's attitude with regard to the dual resonance
model was that he had a great skepticism. 
He thus started with P. Frampton, a postdoc then at Chicago,
to check the positivity of factorized coupling at all resonance poles.
What they found was that except the tachyon
all resonances behave as if they are genuine objects
of positive definite metric.

I believe that this was a turning point to him,
and he started to search for dynamics behind dual resonance models,
ultimately leading to his discovery of the string theory.
Nambu's ideas on string theory are summarized with
a broad perspective in his famous undelivered lecture note
\cite{string}.
When I first saw the original typed manuscript of this lecture note prepared by
Nambu himself,
it had only a description of the geometric construction which
was very beautiful.
The lecture note later presented in the final form gives a broader scope
of string picture in hadron physics, and
this is a good place to learn how Nambu thought,
as explained by P. Ramond at this workshop.

\section
{\bf Symmetry breaking: how they emerge}

One of the greatest contributions of Nambu to particle physics
is the discovery of spontaneous symmetry breaking.
The symmetry idea in particle physics has been popular
since the introduction of isospin by Heisenberg, the hyper-charge
of strangeness due to Nishijima and Gell-Mann and
$SU(3)$ flavor symmetry of Gell-Mann and others.
All these were very useful to classify what were called elementary
particles then, and to relate physical observables by
a selection rule.
Nambu added to this list a new class of symmetry breaking,
or rather a hidden symmetry behind apparent phenomena.
His original application of this general idea was directed to
the $SU(2) \times SU(2)$ chiral symmetry among
nucleons and pions, but it was obvious that the idea
was more general.

The spontaneous symmetry breaking appears since
the symmetry hidden at the fundamental level,
either in the hamiltonian or in the equation of motion,
is not respected by the energy minimum condition
of the ground state.
The conflict between the hidden symmetry
and the spontaneous breaking is restored by
emergence of the long range force,
now called Nambu-Goldstone boson.

Needless to say,
the spontaneous symmetry breaking is more
restrictive than an explicit symmetry breaking
introduced in the case of usual symmetry.
This restriction gave rise to intricacies of
our quantum world.
One of these intricacies is the spontaneous symmetry
breaking of gauged symmetry, known as the Higgs mechanism.
The Nambu-Goldstone mode becomes the longitudinal part
of otherwise purely transverse gauge bosons.
This made the useless Yang-Mills non-Abelian gauge theory 
a leading candidate of particle physics theory.

I suspect that the $U(1)$ Higgs mechanism is well recognized
in Nambu's mind prior to the Higgs paper, due to the Meissner effect and the work
of Anderson.
I am reminded of Nambu's comment on the t' Hooft seminal 
work on the renormalizability of the Weinberg model.
When he mentioned `t Hooft's circulating preprint   to M. Suzuki and myself
then at Berkeley,
he confessed that he had known the work of Weinberg 
since its birth to our great
surprise, because the model was not familiar to many of
particles physicists.

\section
{\bf Yet another symmetry breaking to be discovered}

My quest for the symmetry breaking is somewhat twisted.
I had been much fascinated by 
the simple and the beautiful $SU(5)$ grand unified theory of Georgi and Glashow
ever since its publication. 
Since the baryon number violation predicted by
this model had no experimental support,
I thought that this is a serious defect of the model,
and attempted to recover the baryon number conservation
by imposing some global symmetry \cite{proton stability} .
After this work was finished,
I began to think the other way:
one should  pursue physical consequences of the baryon number 
violation instead of a contrived idea.
This led to my work on baryo-genesis in 1978 \cite{baryo-genesis}.

Incidentally,
I wish to bring to the audience the great contribution
of Hirotaka Sugawara on the birth of neutrino oscillation experiments.
He recognized the importance of my baryo-genesis work with great interest, and
persuaded M. Koshiba to initiate the proton decay experiment
which led to Kamiokande detector.
Although the group was not successful in the search of
proton decay, they later converted the detector
to suit for the solar neutrino search.
With completion of this upgrade,
Koshiba's group succeeded in observing supernova
burst neutrino in the spring of 1987.
Later the group got the money for super-Kamiokande project
which was led by Y. Totsuka.
This upgrade led to this year's Nobel prize for T. Kajita.
Without Sugawara's enthusiasm the story of
neutrino physics in the world would have been different.
I had a fortune to meet Sugawara in my first year of
Chicago where he was spending his postdoctoral year.

It is  my strong belief, which I suspect also shared by many,
that all accidental symmetries not protected by
the gauge principle  are doomed to be broken,
albeit their strength not precisely known.
The baryon and the lepton number conservation valid to
excellent experimental resolution today belong to this class of
symmetry to be broken at some high energy scale.

Fortunately there were some guides such as
$SU(5)$ grand unified theory of Georgi and Glashow 
for the baryon number violation and $SO(10)$ model
for lepton number violation.
The use of lepton number violation for creating
the matter-antimatter imbalance of our universe
became more popular than the baryo-genesis idea,
hence one calls this lepto-genesis.
The popularity of lepto-genesis is understandable,
because one naturally expects that one of
the ingredients for a successful scenario
of the asymmetry generation is already there:
discovery of neutrino oscillation indicates
the finite neutrino mass and its mass
presumably violates the lepton number
according to the idea of Majorana.
On the other hand, there is no experimental
hint on the baryon number violation.

The Majorana nature of neutrino masses
has two attractive features.
We assume that the nature favors a symmetry
of all quarks and leptons,
and that neutral leptons are born with
four components similar to all other fundamental
particles.
On the other hand,
the electroweak interaction described
by the standard theory only requires 
two-component neutrinos, since the right-handed
component is free of quantum numbers of
$SU(3) \times SU(2) \times U(1)$ gauge group.
This suggests that the right-handed component
has a huge mass compared to the electroweak
scale given by the Fermi constant.

The presence of the right-handed component
gives rise to the seesaw mechanism that explains
the smallness of ordinary neutrinos,
caused by a Dirac mass mixing term generated
by the usual Higgs coupling.
Moreover, the right-handed component may
become the agent of lepton asymmetry
generation at an early epoch of cosmological
evolution.
Using the important observation of Shaposhnikov and
others that
lepton and baryon numbers can be converted to
a thermal equilibrium value in high temperature
phase above the electroweak scale,
Fukugita and Yanagida proposed the lepto-genesis
idea which  was further elaborated by many others
later on.

The important prerequisite for a success of
lepto-genesis idea is an experimental proof
of the Majorana nature of ordinary neutrino.
A conventional experimental method of
exploring the Majorana nature of neutrinos
is to look for the lepton number violating
double-beta decay, the neutrino-less
double beta decay.
Although this is an interesting approach to
directly look for the symmetry violation,
the expected tiny neutrino masses below eV
might be a great obstacle against experiments.

\section
{\bf Towards Spectroscopy of Atomic Neutrino (SPAN)}

Our method of looking for the Majorana nature of
neutrinos is quite different, and it
has a merit of evading the smallness of the effect.

Consider atomic de-excitation.
Although tiny, it contains a decay branch into a neutrino
pair  if quantum number changes match.
Its presence is ensured in the standard electroweak theory.
Suppose then that the neutrino is of Majorana type.
Two emitted
neutrinos are then indistinguishable fermions, hence emitted
two-particle state must satisfy the anti-symmetry
under the exchange of the two.
This leads to an interesting interference contribution of rates
where the Majorana mass effects are large.
The effect does not occur for Dirac neutrinos, hence
giving a method of distinction between
Majorana and Dirac neutrinos \cite{my 07 prd}.

Although the method of SPAN (SPectroscopy of Atomic Neutrino)
seems beautiful,
there was a serious problem: the smallness of atomic
processes.
The weak interaction rates scale with energy to high power,
usually to the fifth power, giving a hopeless rate.
Fortunately, we discovered a mechanism of huge enhancement
\cite{macro-coherence initial}, \cite{yst pra}
which I shall describe.

The idea dates back to Dicke's super-radiance (SR).
Dicke proposed that  single photon emission,
when it occurs within the wavelength region
of emitted photon,
may be enhanced.
Instead of the usual stochastic decay
of individual atoms a cooperative phenomena
may occur after some delay:
after several spontaneous emission along
a long target direction  atoms in target acquire a coherent phase
within the wavelength region, and
all of sudden essentially all atoms in the upper level
decay to the lower level.
This is an explosive process and one may
view this as an enhanced decay process.

We discovered \cite{macro-coherence initial}, 
\cite{yst pra}, \cite{renp overview} that this enhancement can be made
much stronger if more than two particles are involved
in the final state.
In Dicke's SR the emitted photon had
a phase factor $e^{i\vec{k}\cdot \vec{x}}$
and this phase may differ at atomic site $\vec{x}$.
The limitation to the wavelength is due to
this different phase at different atomic site.
But if one considers two-photon emission,
the phase factor becomes $e^{i(\vec{k}_1 + \vec{k}_2)\cdot \vec{x}}$, and
at the special phase space point of $\vec{k}_1 + \vec{k}_2 = 0$
there is no restriction to the wavelength.
This condition implies that both the energy and the momentum
are conserved among emitted particle.
(In atomic physics the momentum recoil of atom
is negligibly small and one usually speaks of no
momentum conservation for the usual decay.)

We may call this new type of coherence the macro-coherence.
If atoms maintain the macro-coherence, radiative emission of 
neutrino pair (RENP), $|e \rangle \rightarrow |g\rangle + \gamma + \nu_i \nu_j$,
may also benefit the huge enhancement.
We worked out consequences of the macro-coherence to RENP 
\cite{renp overview}, \cite{dpsty}, \cite{nuclear monopole}.
There are interesting possibilities to explore almost all
important questions left in neutrino physics:
(1) determination of absolute neutrino mass,
(2) distinction of normal vs inverted mass hierarchies,
(3) distinction of Majorana vs Dirac neutrinos,
(4) measurement of three CP violating phases
including the Majorana phase,
(5) possibility of detecting relic neutrino of 1.9 K \cite{relic}.

There are however many experimental problems to overcome
before we write a serious proposal for the neutrino mass
spectroscopy (SPAN).
Some of these problems are already discussed and
I shall not discuss these here.
One thing which was obvious to us
is that the macro-coherence idea may work
for much stronger process than RENP.
We decided to embark on experimental project
of verifying the macro-coherent amplification
in weak 	QED processes.

The process we chose is two-photon emission just mentioned.
The macro-coherently amplified two-photon emission
may be called paired super-radiance (PSR).
In our experiment we used the vibrational transition of
para-hydrogen molecule, pH$_2$, because this process is electric
dipole forbidden and its calculated two-photon decay rate is 
small, $\sim 10^{12}$ sec,
almost impossible to measure in laboratories.
We developed the coherence between two states
of the first excited vibrational state $v=1$ and
the ground state $v=0$ by
irradiation of two Raman-type of lasers
of frequencies $\omega_i, i=1,2$ with $\omega_1 - \omega_2 = 0.52$eV,
the vibrational level spacing.
Two lasers of   blue $\omega_1$ and red  $\omega_2$
frequencies have been irradiated from
the same direction. One expects many side bands
due to multi- photon processes if
the coherence of lasers is good.
These side band photons result from higher order
multiple-photon processes generated by two Raman lasers.
We observed many side bands both
in the blue and the red regions, more than $O(10)$ \cite{psr experiments}.
The existence of many side bands means that
we have achieved a good coherence.
Our estimate of macro-coherence is roughly 
0.08 of the maximum allowed value over
the macroscopic body of 15 cm long.

We observed PSR in two ways.
One of the paired photons is generated by one of the side bands whose frequency
happens to lie between the level interval \cite{psr experiments}.
The other method \cite{psr experiments 2} used an external trigger different from
the side band frequency.
Results of  \cite{psr experiments}, \cite{psr experiments 2} show clear evidences of
PSR phenomena.
The amount of enhancement is estimated more than
$10^{18}$.

The next important step towards SPAN is to form an ideal target state
for RENP.
This is the state in which there is a minimum photon emission from QED processes.
PSR itself may become a serious background, but
there exists an analogue of stopped light extended to two-photon
process in which one expects that photon emission occurs only from
edges of the target and not from the bulk.
If this state is realized, one can minimize the PSR background.
We call this ideal state two-photon soliton  and
its solutions have been found \cite{psr soliton}.

Clearly, there are many experimental and theoretical
works before we write a serious RENP proposal.
We are advancing this step steadily in Okayama.

\section
{\bf Epilogue}

Nambu had  versatile interests in all areas of physics
and basic sciences.
He seemed to convey all the time that physicists should enjoy
scientific achievements using  benefits as
a physicist.
He also showed his interest in people
who created the greatest achievements in science.
I learned lots on these from interaction with Nambu.

Since I left Chicago, I fortunately had many occasions to
talk with him, in particular the chances when 
I could communicate my own works were very
valuable.
When I spoke of the baryo-genesis idea at Sendai in 1978
where he happened to attend the annual meeting
of Japan Physical Society, he showed an immediate
and a great interest in my work.
Later at Osaka I had many chances to talk
about our SPAN project which he encouraged
very much.

I learned from Nambu that
one should be brave in creating and developing
new ideas, but at the same time
one should study physics achievements as a whole.
What I could not learn from Nambu is
to foster ideas for a  very long time over more than ten years.
Some of his bold ideas goes back to his school days.
This is something truly amazing.
He has certainly left to his closest friends great impressions
of how outstanding ideas can be developed.
I hope that the Nambu way is also appreciated by  many.

\end{document}